\documentclass[a4paper,11pt]{article}
\usepackage{jinstpub} % for details on the use of the package, please
\usepackage{arydshln} %for hdashline
\usepackage{amssymb}
\usepackage{amsmath}
\usepackage{threeparttable}
\usepackage{lineno}

\begin{document}

\title{\boldmath A measurement of the scintillation decay time constant of nuclear recoils in liquid xenon with the XMASS-I detector}

\author[a,d]{K.~Abe,}
\author[a,d]{K.~Hiraide,}
\author[a,d]{K.~Ichimura,}
\author[a,d]{Y.~Kishimoto,}
\author[a,d]{K.~Kobayashi,}
\author[a]{M.~Kobayashi,}
\author[a,d]{S.~Moriyama,}
\author[a,d]{M.~Nakahata,}
\author[a,d,1]{H.~Ogawa\note{Now at Department of Physics, College of Science and Technology, Nihon University, Kanda, Chiyoda-ku, Tokyo 101-8308, Japan.},}
\author[a]{K.~Sato,}
\author[a,d]{H.~Sekiya,}
\author[a]{T.~Suzuki,}
\author[a]{O.~Takachio,}
\author[a,d]{A.~Takeda,}
\author[a]{S.~Tasaka,}
\author[a,d]{M.~Yamashita,}
\author[a,d,2]{B.~S.~Yang\note{Now at Center for Axion and Precision Physics Research, Institute for Basic Science, Daejeon 34051, South Korea.},}

\author[b]{N.~Y.~Kim,}
\author[b]{Y.~D.~Kim,}

\author[c,e]{Y.~Itow,}
\author[c]{K.~Kanzawa,}
\author[c]{K.~Masuda,}

\author[d]{K.~Martens,}
\author[d]{Y.~Suzuki,}
\author[d]{B.~D.~Xu,}

\author[f]{K.~Miuchi,}
\author[f]{N.~Oka,}
\author[f,d]{Y.~Takeuchi,}
\author[g,b]{Y.~H.~Kim,}
\author[g]{K.~B.~Lee,}
\author[g]{M.~K.~Lee,}
\author[h]{Y.~Fukuda,}
\author[i]{M.~Miyasaka,}
\author[i]{K.~Nishijima,}

\author[j]{K.~Fushimi,}
\author[j]{G.~Kanzaki,}

\author[k]{S.~Nakamura,}

\begin{center}
\collaboration[c]{XMASS Collaboration}
\end{center}

\affiliation[a]{Kamioka Observatory, Institute for Cosmic Ray Research, the University of Tokyo, Higashi-Mozumi, Kamioka, Hida, Gifu, 506-1205, Japan}
\affiliation[b]{Center for Underground Physics, Institute for Basic Science, 70 Yuseong-daero 1689-gil, Yuseong-gu, Daejeon, 305-811, South Korea}
\affiliation[c]{Institute for Space-Earth Environmental Research, Nagoya University, Nagoya, Aichi 464-8601, Japan}

\affiliation[d]{Kavli Institute for the Physics and Mathematics of the Universe (WPI), the University of Tokyo, Kashiwa, Chiba, 277-8582, Japan}
\affiliation[e]{Kobayashi-Maskawa Institute for the Origin of Particles and the Universe, Nagoya University, Furo-cho, Chikusa-ku, Nagoya, Aichi, 464-8602, Japan}
\affiliation[f]{Department of Physics, Kobe University, Kobe, Hyogo 657-8501, Japan}
\affiliation[g]{Korea Research Institute of Standards and Science, Daejeon 305-340, South Korea}
\affiliation[h]{Department of Physics, Miyagi University of Education, Sendai, Miyagi 980-0845, Japan}
\affiliation[i]{Department of Physics, Tokai University, Hiratsuka, Kanagawa 259-1292, Japan}
\affiliation[j]{Department of Physics, Tokushima University, 2-1 Minami Josanjimacho Tokushima city, Tokushima, 770-8506, Japan}
\affiliation[k]{Department of Physics, Faculty of Engineering, Yokohama National University, Yokohama, Kanagawa 240-8501, Japan}

\emailAdd{xmass.publications12@km.icrr.u-tokyo.ac.jp}

\abstract{We report an in-situ measurement of the nuclear recoil (NR) scintillation decay time constant in liquid xenon (LXe) using the  XMASS-I detector at the Kamioka underground laboratory in Japan.
XMASS-I is a large single-phase LXe scintillation detector whose purpose is the direct detection of dark matter via NR which can be induced by collisions between Weakly Interacting Massive Particles (WIMPs) and a xenon nucleus.  The inner detector volume contains 832 kg of LXe. 

 $^{252}$Cf was used as an external neutron source for irradiating the detector.  The scintillation decay time constant of the resulting neutron induced NR was evaluated by comparing the observed photon detection times with Monte Carlo simulations.  
Fits to the decay time prefer two decay time components, one for each of the Xe$_{2}^{*}$ singlet and triplet states, with $\tau_{S}$ = 4.3$\pm$0.6 ns taken from prior research, $\tau_{T}$ was  measured to be 26.9$^{+0.7}_{-1.1}$ ns with a singlet state fraction F$_{S}$ of 0.252$^{+0.027}_{-0.019}$.
We also evaluated  the performance of pulse shape discrimination  between NR and electron recoil (ER) with the aim of reducing the electromagnetic background in WIMP searches.  For a 50\% NR acceptance, the ER acceptance was 13.7${\pm}$1.0\% and 4.1${\pm}$0.7\% in the energy ranges of  5--10 keV$_{\rm ee}$ and  10--15 keV$_{\rm ee}$, respectively.}


\arxivnumber{1234.56789} % only if you have one

\maketitle
%\begin{linenumbers}

%\flushbottom
\section{Introduction}
\label{sec:intro}
Liquid xenon (LXe) has been used in many modern experiments such as dark matter  and neutrino-less double beta decay searches~\cite{Abe:2013tc,Akerib:2012ys,Cao:2014jsa,Aprile:2017aty,Auger:2012gs}. 
The scintillation timing information can be used for  position reconstruction of an event in the detector~\cite{XMASS:2018bid} as well as  for  particle identification~\cite{Abe:2018gyq}.
Studies on the scintillation process in LXe has been conducted in various experiments~\cite{Takiya:2016wio,Kubota1978,Hitachi:1983zz,Kubota1979,Keto1979,Akimov:2001pb,Dawson:2005kb,Teymourian:2011rs,Murayama:2014wwa,UeshimaD,Akerib:2018kjf,Hogenbirk:2018zwf}.
A scintillation photon is produced by two mechanisms. One is the direct excitation of Xe atoms that then forms an excited dimer Xe$_{2}^{*}$,
 \begin{eqnarray*}
\mathrm{Xe}^{*} + \mathrm{Xe} + \mathrm{Xe} &\rightarrow&  \mathrm{Xe}_{2}^{*} + \mathrm{Xe}~, \\
\mathrm{Xe}_{2}^{*}&\rightarrow& 2\mathrm{Xe} + h\nu ~.  \\
\end{eqnarray*} 
The other process involves  the recombination process between electrons and Xe ions
\begin{eqnarray*}
\mathrm{Xe}^{+} + \mathrm{Xe} &\rightarrow&  \mathrm{Xe}_{2}^{+}~, \\
\mathrm{Xe}_{2}^{+} + \mathrm{e}^{-}&\rightarrow& \mathrm{Xe}^{**} + \mathrm{Xe}~,  \\
 \mathrm{Xe}^{**}&\rightarrow& \mathrm{Xe}^{*} + heat~, \\
  \mathrm{Xe}^{*} + \mathrm{Xe} + \mathrm{Xe} &\rightarrow&  \mathrm{Xe}_{2}^{*} + \mathrm{Xe}~,\\
  \mathrm{Xe}_{2}^{*}&\rightarrow& 2\mathrm{Xe} + h\nu~.  
\end{eqnarray*} 
The Xe$_{2}^{*}$ dimer has both a singlet and triplet state, each with its own decay time constant. 
The decay time constants of the singlet and triplet states were reported to be 4.3$\pm$0.5  ns and 21$\pm$2 ns, respectively using $^{252}$Cf fission fragments~\cite{Hitachi:1983zz}. While the recombination process has a longer decay time constant of more than 30 ns, measured with 1 MeV electrons from a $^{207}$Bi  source~\cite{Hitachi:1983zz,Kubota1979}. 
In the case of neutron induced NR events, the decay time constant of the triplet state was reported to be $\sim$20 ns both with an applied electric field  (0.1--0.5 kV/cm)~\cite{Akerib:2018kjf,Hogenbirk:2018zwf} and without~\cite{Akimov:2001pb,UeshimaD}.
The decay time constants of the singlet and triplet states  depend weakly on the density of the excited species, whereas the  ratio of the singlet to triplet state as well as the recombination time depends on the deposited energy density~\cite{Aprile:2009dv}. 
The time profile of events' scintillation photon hits (pulse shape) may allow for discrimination between NR and ER initiated events~\cite{Ueshima:2011ft}.

XMASS-I  is a large single phase LXe detector, built primarily for dark matter searches, previously reported the Xe$_{2}^{*}$ triplet decay time constant of ER events using low energy gamma-rays calibration sources~\cite{Takiya:2016wio}.
In this work,  we  measured the Xe$_{2}^{*}$ triplet decay time constant of NR events using an external $^{252}$Cf neutron source irradiating the XMASS-I detector and also evaluated the usefulness of  pulse shape discrimination (PSD) between NR and ER in the energy region of interest for  dark matter searches.

\section{Experimental apparatus}
\subsection{The XMASS-I detector}
The XMASS-I  detector is located in the Kamioka mine under 1,000 m of rock (2,700 meter water equivalent).
As shown in Fig.~\ref{fig:NeutronCalibrationPipe}, the inner detector (ID) contains 832 kg of LXe inside a spherical, oxygen free high conductivity (OFHC) copper structure with an 80 cm diameter.  Scintillation light from the LXe is detected by 630 hexagonal R10789 photomultiplier tubes (PMTs)  and 12 cylindrical R10789Mod PMTs with a total photocathode coverage of 62.4\%. 
 The inner containment vessel contains the  LXe and the PMT holder, while  the outer containment vessel holds  vacuum for thermal insulation.  In order to reduce  external gamma-rays and neutrons from the surrounding rock, the ID is placed at the center of the outer detector (OD). The OD is  a cylindrical  tank 10 m in diameter and 11 m in height  filled with ultrapure water. 72 Hamamatsu  20-inch R3600 PMTs  are mounted on the inner surface of the water tank to provide an active muon veto. More details  can be found in Ref.~\cite{Abe:2013tc}.

Signals from the 642 ID PMTs were recorded by CAEN  V1751 waveform digitizers with a 1 GHz sampling rate and 10-bit resolution.  Analog-timing-modules (ATMs) that were previously used in the Super-Kamiokande experiment~\cite{Ikeda:1992cy,Fukuda:2002uc}  worked for generating a trigger.   The threshold for an ID PMT  to register a hit in the ATMs is set at 0.2 photoelectron (PE). When 4 or more ID PMT hits are observed in a 200 ns coincidence window, a global trigger is issued to both the ATMs and the waveform digitizers. For each triggered event, the waveform is recorded with a width of 10 $\mu$s. The OD trigger requires at least 8 PMT hits in a 200 ns coincidence window.

\subsection{Detector calibrations}
\subsubsection{LED calibration}
The individual PMT gains are monitored by a blue LED embedded in the inner surface of the PMT holder. 
This LED is flashed every second  using the one-pulse-per-second signal from the global positioning system.  LED calibration data is taken continually  during the physics runs and identified by  the trigger  information.

\subsubsection{Energy calibration and light yield}
To check the stability of the detector's light yield,  inner calibration data using a $^{57}$Co source is taken every one or two weeks. Deploying the $^{57}$Co 122 keV gamma-ray source at the center of the detector, we obtain a photoelectron yield of $\sim$15 PE / keV and also trace the timing offsets of the PMT channels.
%In addition, the $^{55}$Fe, $^{109}$Cd, $^{241}$Am, and $^{137}$Cs sources  are also used for the position-dependent energy calibration of the detector.
The $^{57}$Co source as well as $^{55}$Fe, $^{109}$Cd, $^{241}$Am, and $^{137}$Cs are also measured off of center along the detector's $z$-axis for position dependent energy calibration within the detector.

\subsection{The neutron source and  its deployment}
 $^{252}$Cf undergoes  spontaneous fission with a branching ratio of 3.11\%. An average fission event emits 8 gamma-rays with a total energy of 7 MeV and 3.75 neutrons~\cite{TEValentine2001}. This gamma-ray emission is used to tag such fission events.
\begin{figure}\centering
\includegraphics[width=12cm]{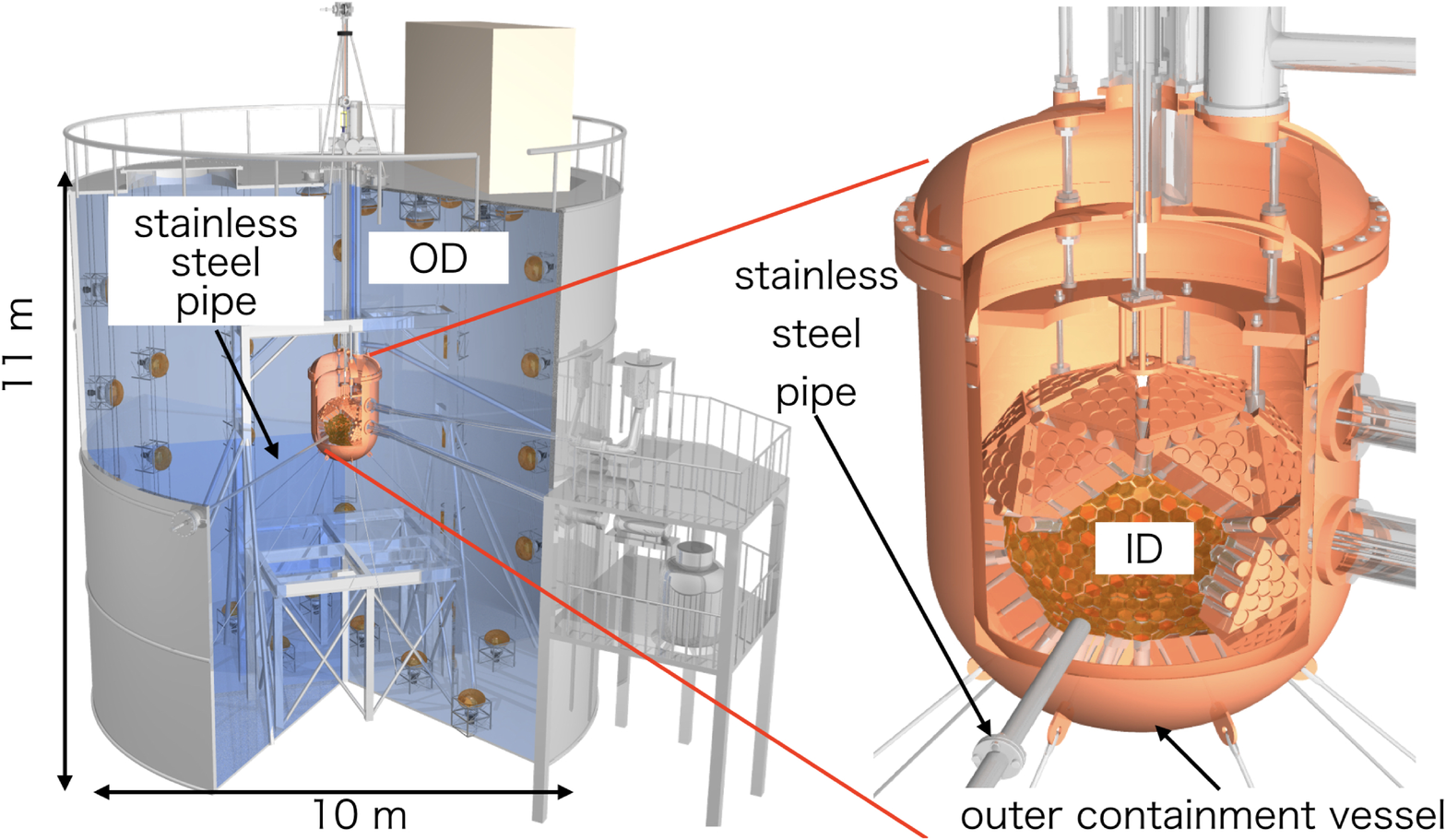}
\includegraphics[width=12cm]{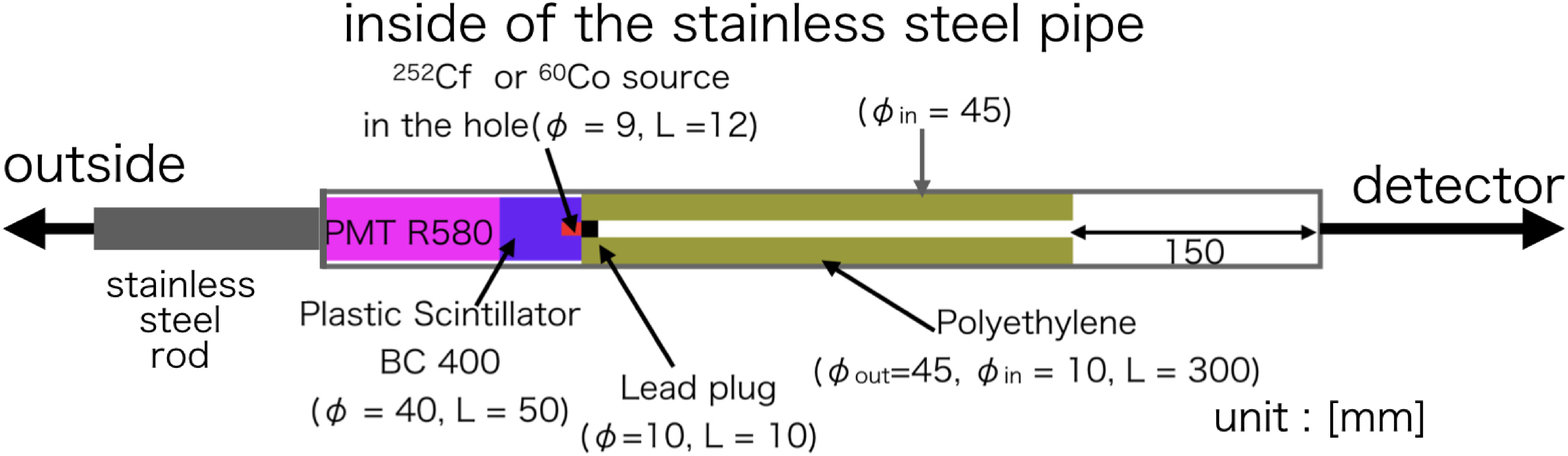}
\caption{Schematic view of the XMASS detector and the calibration setup. }
\label{fig:NeutronCalibrationPipe}
\end{figure}
Figure~\ref{fig:NeutronCalibrationPipe}  shows the  calibration setup.
To detect the gamma-rays, the $^{252}$Cf source was deployed in the cylindrical Bicron BC400 plastic scintillator, which is 40 mm in diameter, and 50 mm long. 
% It has a central hole 9 mm in diameter and 12 mm deep to place the source.
 It has a central hole with  a diameter of 9mm and  a depth of 12 mm, in which the source is placed.
The plastic scintillator was coupled with a Hamamatsu R580 1.5 inch PMT, and hereafter we call it  a neutron tagging assembly (NTA).
A timing calibration between the plastic scintillator and XMASS-I detector electronics was performed with the $^{60}$Co source. The signal of the plastic scintillator was recorded by the same waveform digitizer used for the ID signals. 
A cylindrical polyethylene pipe with the outer diameter of 45 mm, the inner diameter of 10 mm, and the length of 300 mm was installed in front of the plastic scintillator and worked as a support structure. A 10 mm long lead plug  is used to shield gamma-rays from the source.
%A cylindrical polyethylene pipe, with an outer diameter of 45 mm, inner diameter of 10 mm, and 300 mm long, was positioned against the plastic scintillator.
%It not only acts as a support structure but also houses a 10 mm long lead plug, face to face with the source, to prevent $^{252}$Cf gamma-rays from reaching the detector.
A stainless steel pipe, with a 45mm inner diameter passes through the water tank and terminates at 10 mm away from the outer containment vessel.
This NTA was inserted into a stainless steel pipe.  A stainless steel rod (SUS rod) attached to the NTA can position the NTA 150 mm away from the end of the stainless steel pipe. This distance is set so the detector trigger rate does not exceed 100 Hz; capable rate of the data acquisition (DAQ).

\section{Analysis}
\subsection{Monte Carlo simulations}
The XMASS Monte Carlo (MC) simulation  is based on Geant4~\cite{Agostinelli:2002hh}.  It includes a detailed detector geometry, particle tracking, the Xe scintillation process, photon tracking, PMT response, and responses from the electronics.  
%For the input parameters of the simulation, the optical parameters, such as the absorption and scattering lengths of LXe, are extracted from the comparison between data and simulation for the $^{57}$Co calibrations with different source positions.  
Table \ref{tbl:MCSummary}  summarises the input parameters used in the MC simulation. The optical parameters, such as the absorption and scattering length of LXe, are extracted by the comparison between data and MC simulation of the $^{57}$Co calibrations at multiple source positions. Gamma-ray events originating close to the inner detector surface situated between PMTs was used to deduce the copper reflection by comparing the PE spectrum from observed data and MC simulation~\cite{XMASS:2018bid}. 
%Table \ref{tbl:MCSummary}  summarizes the input parameters of the MC simulation.
%The copper reflection was deduced from the comparison of gamma-rays from the surface of the detector between data and MC simulations. 

\begin{table*}[htbp]
\centering
\caption{Summary of input parameters used in the MC simulation. These were obtained by comparing observed data and MC simulations for a range of calibration techniques outlined above. }
\smallskip
\begin{tabular}{|l|c|}
\hline
LXe density & 2.89 g/cm$^{3}$ \\

Emission spectrum of LXe 
&  Gaussian distribution (Ref.~\cite{Fujii2015} )	\\

& centered at 174.8 nm, FWHM 10.2 nm \\

Refractive index of LXe
& 1.58--1.72 for 183--167 nm	\\
 &  from direct measurements~\cite{Nakamura2007} with a small correction \\
 & considering density dependence \\ 
 
LXe absorption length
& 852.9 cm\\

LXe scattering length
& 52.7 cm \\

PMT average quantum efficiency 
& 30\% \\

PMT quartz absorption length 
& 14.3 cm at 175 nm, measured by manufacturer  \\

Copper  reflectivity
& 25\% \\

\hline 
\end{tabular}
\label{tbl:MCSummary}
\end{table*}	

In the MC simulation of the $^{252}$Cf calibration, the Brunson model~\cite{TEValentine2001}  and 
the Watt spectral model~\cite{Frohner1990}  were used for the input energy spectrum of the gamma-rays and the neutrons, respectively. 
A neutron can either be captured by a xenon nucleus or simply interact with it via elastic or inelastic scattering. The cross sections from both the ENDF/B-VII.0 library, and the G4NDL3.13 library based on the ENDF/B-VI library were used and the results compared in order to evaluate the cross section's systematic uncertainty.
We followed the instructions in Ref.~\cite{EMendoza2012,EMendoza2014} to use ENDF/B-VII.0 library.
%For the cross section of neutron elastic, and inelastic scattering off xenon nucleus and neutron capture, we used both the ENDF/B-VII.0 library following the instructions in Ref. \cite{EMendoza2012,EMendoza2014} and the G4NDL3.13 library based on the ENDF/B-VI library, and compared the results to evaluate the systematic uncertainty of the cross section.  
In considering NR events, the relative scintillation efficiency  L$_{\rm {eff}}$~\cite{Aprile:2011hi}, is defined as the scintillation yield of xenon for NR relative to the zero-field scintillation yield for 122 keV gamma-rays from $^{57}$Co. The non-linearity of scintillation yield of ER events over energy was accounted for using a model from Ref.~\cite{Doke2003} tuned with XMASS-I gamma-ray calibration data.
%For NR events, we considered the relative scintillation efficiency, named  L$_{\rm {eff}}$ \cite{Aprile:2011hi}, which is defined as the scintillation yield of xenon for NR relative to the zero-field scintillation yield  for 122 keV gamma-rays from $^{57}$Co. As for ER events, the non-linearity of scintillation yield was taken into account using a model from Ref.  \cite{Doke2003} tuned with XMASS-I gamma-ray calibration data. 
Using the measurement setup outlined in Fig.~\ref{fig:NeutronCalibrationPipe}, the rate for multiple neutrons or gamma-rays from the same fission event to enter the ID simultaneously was found to be negligible. Therefore, only an individual neutron or gamma-ray was generated for each MC event in the $^{252}$Cf simulation with their intensities considered.
%Since the rate for multiple neutrons or gamma-rays from the same fission event to enter the ID simultaneously was found to be negligible using this measurement setup, we generated a single neutron or a gamma-ray for each MC event in the $^{252}$Cf simulation while considering their intensity.

The detection time of the $i^{\rm th}$ photon $T^{i}$ after a $^{252}$Cf spontaneous fission is  defined as
\begin{equation}
\label{eq:MCTime1}
T ^i= t_{\rm Edep} + t^{i}_{\rm scinti} + t^{i}_{\rm TOF} +t^{i}_{\rm TT}+ t^{i}_{\rm jitter}.
\end{equation}
 $t_{\rm Edep}$ is the time when the incident particle deposits its energy in the LXe. And the LXe scintillation photon emission time $t^{i}_{\rm scinti}$ follows the scintillation decay time profile parameterized as 
\begin{equation}
\label{eq:MCTime2}
f(t) = \frac{F_{S}}{\tau_{S}}\cdot \mathrm{exp}\left(-\frac{t}{\tau_{S}}\right) + \frac{1 - F_{S}}{\tau_{T}}\cdot \mathrm{exp}\left(-\frac{t}{\tau_{T}}\right).
\end{equation}

We assumed that the scintillation decay time profile has two decay constants  $\tau_{S}$  and $\tau_{T}$, corresponding to the decay constants of the singlet and the triplet states respectively, and that the fraction of photons generated from decays of singlet dimers $F_{S}$ and triplet dimers $F_{T}$ sum to unity ($F_{T}$ = 1 - $F_{S}$)  following Ref.~\cite{Takiya:2016wio}. 
%A possible third component owing to the recombination process, contributes at most 10\% of total scintillation light for nuclear recoil \cite{Aprile:2006kx} and is not considered here.
%We assumed that the scintillation decay time constant has two components $\tau_{S}$  and $\tau_{T}$, corresponding to the decay constants of the singlet and the triplet state following Ref. \cite{Takiya:2016wio}, respectively. The recombination process contributes at most 10\% of total scintillation light for nuclear recoil \cite{Aprile:2006kx}.  
%$F_{S}$ denotes the fraction of photons generated from decays of singlet dimers.
 $ t^{i}_{\rm TOF}$ is the time of flight (TOF) of the scintillation photon. 
Here, the group velocity of the scintillation light was calculated from  the refractive index of LXe.
 $t^{i}_{\rm TT}$ is the transit time in the  PMT, which we assume to be the same  for all PMTs. 
The transit time spread (TTS) of  $\sigma$ = 2.4 ns  for  PMT \cite{Uchida:2014cnn} was included in the timing calculation.
$t^{i}_{\rm jitter}$ is a  smearing parameter accounting for the timing jitter in the electronic channel of PMT and extracted from the $^{57}$Co calibration data. 
 It follows a Gaussian distribution with a standard deviation of  0.93 ns \cite{Takiya:2016wio}.
%$t^{i}_{\rm jitter}$ is a  smearing parameter to reproduce the uncertainty of the electronics simulation. It follows a Gaussian distribution with a standard deviation of 0.93 ns extracted from the $^{57}$Co calibration data\cite{Takiya:2016wio}. 
After calculating $T^{i}$ for all photons, a waveform for each PMT was simulated  using the one PE pulse (template pulse) shape extracted from LED calibration data.  
%The integrated template pulse represents the one PE of real data at the level of $\sigma$ < 0.1 PE.
A residual between the template pulse and the real data was found to be $\sigma$ < 0.1 PE for all PMTs.
\subsection{Event selection}
\label{sec:EventSelection}
\begin{figure}\centering
\includegraphics[width=10cm]{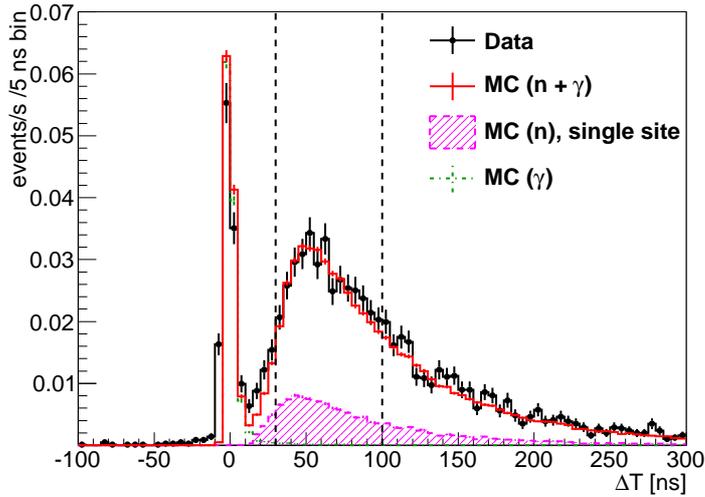}

\caption{Timing distributions of $^{252}$Cf source data (solid black) and MC simulation (solid red).  $\Delta T$ is defined in section \ref{sec:EventSelection}.   The timing of each event is aligned so that the events induced by the $^{252}$Cf gamma-rays peak at $\Delta T$ = 0 ns. Magenta dotted and green dash-dotted histograms show the timing distribution from the  $^{252}$Cf neutron simulation (single site events only) and the $^{252}$Cf gamma-ray simulation, respectively.  The accidental coincidence contribution was derived from the event rate in -1900 $<$ $\Delta T$ $<$ -100 ns time window. This accidental  rate was about 3 $\times$ 10$^{-4}$ events/s and was subtracted from data. Events in 30 < $\Delta T$ < 100 ns were used for the NR events analysis.}
\label{fig:TOF}
\end{figure}

We took 1.5 hours of $^{252}$Cf source data  with an ID trigger rate of roughly 80 Hz.
The signal of the plastic scintillator was searched for by offline analysis. 
Events related to neutrons from the $^{252}$Cf source were selected using the following three criteria. 
(1) Only the ID trigger is issued to avoid muon or muon induced events.  (2) The time difference from the previous ID event is longer than 500 $\mu$s and the root mean square of the timings of all hits in the event is less than 100 ns. This cut removes noise events that often follow particularly high energy events. (3) ID trigger is issued between 30 and 100 ns after the NTA trigger  ($\Delta T$)  to avoid $^{252}$Cf gamma-rays induced events.  
%As the response of the NTA is not implemented in  MC simulation, $\Delta T$ in MC simulation is calculated from the time difference between the time of the particle generation and the time that the ID trigger issued.
$\Delta T$ in MC simulation is calculated from the time difference between  particle generation and an ID trigger being issued. 
Figure~\ref{fig:TOF} shows the $\Delta T$ distributions of events which have less than 500 detected PE in the ID. The neutron in a MC event can deposit its energy at a single position or over multiple positions within the LXe. If all of the photons detected originate from a single position they are classed as single-site, if multiple positions then multi-site. 

\begin{figure}\centering
\includegraphics[width=10cm]{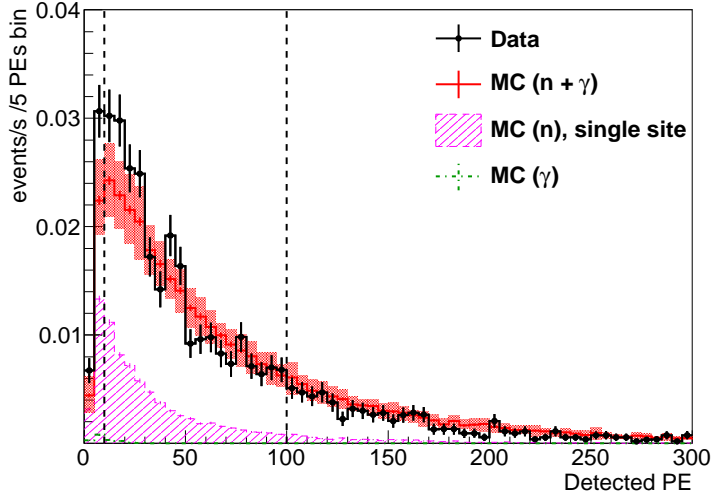}
\caption{Detected PE distributions of $^{252}$Cf source data (solid black) and MC simulation (solid red). The red band shows the quadratic sum of the systematic uncertainty and the statistical uncertainty of the MC simulation. The magenta hashed histogram and the green dash-dotted histogram show the single-site events from the $^{252}$Cf neutron simulation, $^{252}$Cf gamma-rays simulation, respectively. The contribution of the accidental coincidence events was derived from events  in -1900 $<$ $\Delta T$ $<$ -100 ns time window and was subtracted from data. Events with PE counts between  10 and 100 PEs were used for NR events analysis.}
\label{fig:Energy}
\end{figure}

Figure~\ref{fig:Energy} shows the energy spectrum after all cuts.  It includes the systematic uncertainties related to the detection efficiency of the plastic scintillator ($\pm$10\%), the cross section difference between the G4NDL3.13 library and the ENDF-B/VII.0 library (25\% at most), the scintillation efficiency for NR ($\pm$1$\sigma$ in Ref.~\cite{Aprile:2011hi}, 10\% at most).
%From the $^{57}$Co calibration data and MC with different source positions, we observed that the mean value of observed PE discrepancy was at most 5\% at the larger radius ( >40 cm).  Therefore, this was also taken into account as a systematic uncertainty in MC simulation in Fig. \ref{fig:Energy}.
We observed a discrepancy (5\% at most) in the mean observed PE between the $^{57}$Co calibration data and MC simulation at large radii (>40 cm). This was included as a systematic uncertainty in the MC simulation in Fig.~\ref{fig:Energy}.
The detection efficiency and tagging threshold of the plastic scintillator were estimated to be (70$\pm$10)\% and 100 keV in gamma-ray energy, respectively,  by comparing the data from   $^{137}$Cs and $^{60}$Co to  MC simulations using a small setup. 
%The detection efficiency of the plastic scintillator was estimated to be (70$\pm$10)\% with a tagging threshold of 100 keV by comparing the count rate for the  $^{137}$Cs and $^{60}$Co source between data and MC simulation using a small setup. 
%We also checked the energy threshold of the plastic scintillator with 10 kBq $^{60}$Co source. We took data with the plastic scintillator  and we also simulated with this geometry. MC with tagging threshold of 100 keV has same tagging efficiency with that of data.
The $\Delta T$ distribution of the MC simulation agreed with data as shown in Fig.~\ref{fig:TOF}.
There was about 25\% count rate difference below 20 detected PEs in the energy spectra in Fig.~\ref{fig:Energy}, we discuss its impact on the decay time constant  in section \ref{sec:NRDecayTime}. 

Multi-site events make up about 75\% of the neutron events as deduced from MC simulation.
%The contribution of the multi-site events is about 75\% of neutron events as deduced from MC simulation.
 Events with PE counts between  10 and 100 PEs, corresponding to the energy range from 1.5 keV$_{\rm ee}$ (6.3 keV$_{\rm nr}$) to 8.3 keV$_{\rm ee}$ (40 keV$_{\rm nr}$), were used to evaluate the NR decay constant described in the following section.

\subsection{Evaluation of the nuclear recoil decay time constant}
\label{sec:NRevaluation}	
\begin{figure}\centering
\includegraphics[width=10cm]{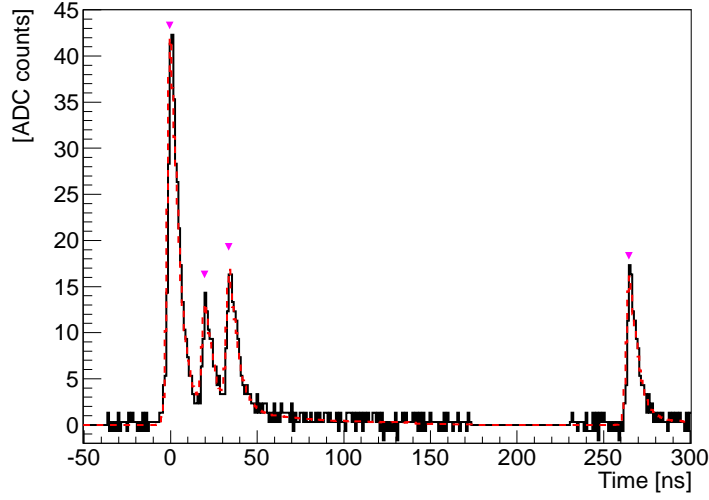}
\caption{A  raw waveform in a single PMT in a $^{252}$Cf calibration event. To reduce data size, parts of a waveform that do not exceed a threshold of 3 ADC counts were not recorded. The magenta triangles show the detected peaks. The dashed red line shows the  waveform reconstruction obtained as the sum of each peak's single-PE waveform fit.  A typical single-PE pulse has a height of about 22.4 ADC counts after digitization.}
\label{fig:WF}
\end{figure}

The scintillation decay time constant of NR was evaluated by comparing the time-distributions of the detected PE over all PMTs and events between data and MC simulation with various timing parameters. 
To analyze waveforms of individual PMTs, we developed a peak finding algorithm based on a Savitzky-Golay filter \cite{ASavitzky1964} to obtain individual photon hit timings. Each peak was fitted with a single PE waveform template obtained from the LED calibration data.  Figure~\ref{fig:WF} shows a NR event waveform from a single PMT with the fitting result. Due to  fluctuations of the baseline and electronic noise, the peak-finding algorithm sometimes misidentifies the tail of the single PE distribution as a peak. Such misidentified peaks typically have  PE smaller than 0.5 PE.  In this study, only peaks that have more  than 0.5 PE are used. 
For each event, all peaks from all PMTs  are sorted in order of detected timings. The timing of the fourth earliest peak is set to $T$ = 0 ns with all other peak timings within the event shifted relative to this time, reflecting the trigger implementation in DAQ. This allows to superimpose all the recorded peak times over events.
%For each event, all peaks from all PMTs are sorted in order of detected timings, and the timing of the fourth earliest peak is set to $T$ = 0 ns, reflecting  the trigger implementation in DAQ.  All peak timings in an event are shifted relative to the time of the fourth earliest peak, to sum up all events.

To obtain the scintillation decay time constant  for NR, we performed  a $\chi^{2}$ fit defined as 
\begin{equation}
\label{eq:Chi2Definition}
\chi^{2} = \sum_{i}\frac{(N_{i}^{\rm data} - N_{i}^{\rm MC} \cdot S)^{2}}{\sigma_{\rm stat(\rm data)}^{2} + \sigma_{\rm stat(\rm MC)}^{2}\cdot S^{2}}~,
\end{equation}
where $N_{i}^{\rm data}$ and $N_{i}^{\rm MC}$ are the number of detected peaks in the $i^{\rm th}$ time bin  over all of the data and simulated MC,  respectively. $S$ is a free variable for normalization.   $\sigma_{\rm stat(\rm Data)}$ and $\sigma_{\rm stat(\rm MC)}$  represent the statistical uncertainty in the data sample and the MC simulation, respectively.

The time bin width was 1 ns and the $\chi^{2}$ is calculated in the range of 3 $\leq T \leq$ 120 ns.
To evaluate the scintillation decay time constant for  NR,
we scanned the parameter $F_{S}$ from 0.0 to 0.5 in steps of 0.025, and $\tau_{T}$ from 21.0 to 30.5 ns in steps of 0.5 ns in  MC simulation.
For $\tau_{S}$, we used 3.7, 4.3, and 4.9 ns taken from Ref.~\cite{Hitachi:1983zz}.

\section{Result and discussion}
\subsection{The scintillation decay time constant of the nuclear recoil}\label{sec:NRDecayTime}

\begin{figure}\centering
\includegraphics[width=10cm]{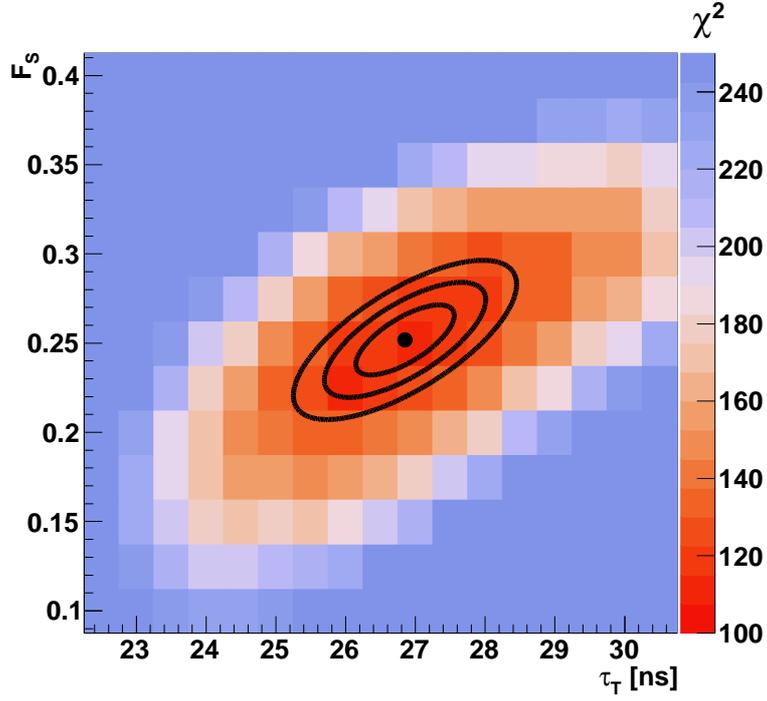}
\caption{ $\chi^{2}$ map in the $F_{S}$ and $\tau_{T}$ plane with $\tau_{S}$ = 4.3 ns.  Black contours correspond to 1$\sigma$, 2$\sigma$, and 3$\sigma$ and are derived from a parabolic function fit described in Eq. \ref{eq:parabolafit}. The black point of (26.9 ns, 0.252) for ($\tau_{T}$,$F_{S}$) is the best fit value from the parabolic function fit.}
\label{fig:Chi2}
\end{figure}

Figure~\ref{fig:Chi2} shows the $\chi^{2}$ map in the $F_{S}$--$\tau_{T}$ plane.  Since the parameters were scanned in discrete steps, a parabolic function fit as defined in Eq. \ref{eq:parabolafit} was performed to obtain the scintillation decay time constant.

\begin{figure}\centering
\includegraphics[width=10cm]{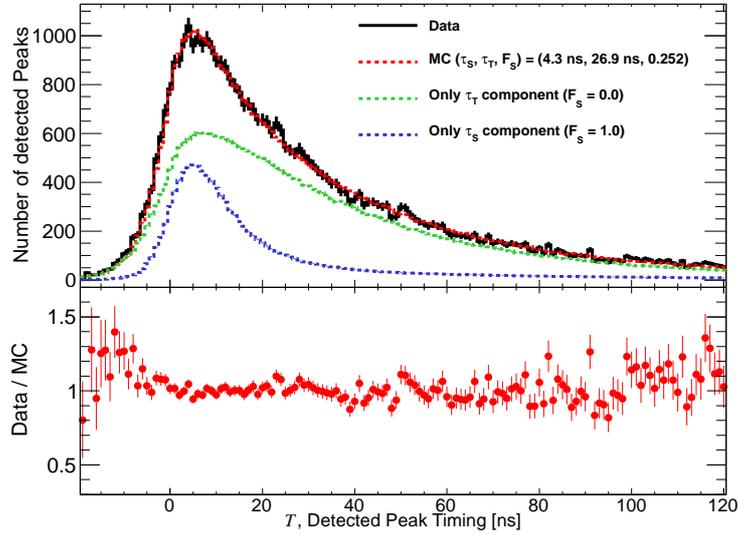}
\caption{(Top) Pulse timing distributions of  data (black) and the best fit MC simulation (red) ($\tau_{S}$,$\tau_{T}$,$F_{S}$) = (4.3 ns, 26.9 ns, 0.252). This MC distribution's two components are also shown separately in green ($\tau_{T}$) and blue ($\tau_{S}$). (Bottom) The ratio of data to the best fit MC simulation. }
\label{fig:FitResult}
\end{figure}

\begin{equation} 
\label{eq:parabolafit}
\begin{split}
\chi^{2} & = a\cdot(\tau_{T}^{\prime})^{2} + b\cdot(F_{S}^{\prime})^{2} + \chi^{2}_{\rm min}~, {\rm where} \\
       \tau_{T}^{\prime} &= (\tau_{T} - \tau_{T}^{\rm best})\cdot {\rm cos}\theta + (F_{S}-F_{S}^{\rm best})\cdot {\rm sin}\theta~ {\rm and}\\
       F_{S}^{\prime} &= -(\tau_{T} - \tau_{T}^{\rm best})\cdot {\rm sin}\theta + (F_{S}-F_{S}^{\rm best})\cdot {\rm cos}\theta~. 
\end{split}
\end{equation}

Here $a$, $b$, $\chi^{2}_{\rm min}$, $\theta$, $\tau_{T}^{\rm best}$ and $F_{S}^{\rm best}$ are free parameters in the fit.
The section of the parameter plane that has  $\chi^{2}$ - $\chi^{2}_{\rm min}$ $<$ 20 in Fig.~\ref{fig:Chi2} was used for this  fit.
From this parabolic fit,  ($\tau_{S}$, $\tau_{T}$, $F_{S}$) were  discovered to be (4.3 ns, 26.9$\pm$0.5  ns (stat), 0.252$\pm$0.013 (stat)) with the minimum $\chi^{2}$ / ndf = 113.9 / 115. 
Data overlaid with the best fit MC simulation  is shown in Fig.~\ref{fig:FitResult}.

All  systematic uncertainties are listed in Table \ref{tbl:ErrorSummary}. They were evaluated as follows:

\begin{table*}[htbp]
\centering 

\caption{Summary of systematic uncertainties on $\tau_{T}$ and $F_{S}$.}
\smallskip
\begin{tabular}{| l | c | c | }
\hline 
Error source & $\sigma_{\tau_{T}}$	(ns) &$\sigma_{F_{S}}$\\
\hline
$\tau_{S} $   &	+0.5, -0.2	& +0.009, -0.003\\

Neutron cross section
&	+0.2, -0.0 & +0.006, -0.000\\
L$_{\rm eff}$
& +0.0, -0.7 & +0.019, -0.010\\

Jitter
& +0.1, -0.2 &	 +0.009, -0.010\\
 After/Pre-pulse
& +0.0, -0.6 & +0.000, -0.001\\
\hline 
Total	&
+0.5, -1.0& +0.024, -0.014 \\
\hline
\end{tabular}
\label{tbl:ErrorSummary}
\end{table*}	

\begin{description}
\item (1)$\tau_{S}$:  %$\tau_{S}$ = 4.3$\pm$0.6 ns from Ref.  \cite{Hitachi:1983zz}  was used in this analysis.  Parameter scans of $F_{S}$ and $\tau_{T}$ with $\tau_{S}$ of 3.7 and 4.9 ns were conducted, and the difference of the respective best fit value was used  as the systematic uncertainty.
%Ref. \cite{Hitachi:1983zz} reported $\tau_{S}$ = 4.3$\pm$0.6 ns using fission fragments from $^{252}$Cf source.
%Ref. \cite{Hitachi:1983zz} reported $\tau_{S}$ = 4.3$\pm$0.6 ns using fission fragments from $^{252}$Cf source.
$\tau_{S}$ = 4.3$\pm$0.6 ns from Ref.~\cite{Hitachi:1983zz}  was used in this analysis.
The systematic uncertainty introduced by $\tau_{S}$ was evaluated by changing $\tau_{S}$  from 4.3 ns to 3.7 ns and 4.9 ns.
%Parameter scans of $F_{S}$ and $\tau_{T}$ with $\tau_{S}$ of 3.7 and 4.9 ns were conducted, and the difference of the respective best fit value was used  as the systematic uncertainty 
%The systematic uncertainty introduced by $\tau_{S}$ was evaluated by changing $\tau_{S}$. 

\item (2){\it Neutron cross section}:  the event rate, including the fraction  of multi-site events, depends on the neutron cross-section.
We found an event rate difference of about 10\% between the MC simulation using the ENDF-B/VII.0  and the MC simulation using the G4NDL3.13 library. Parameter scans of $F_{S}$ and $\tau_{T}$ with the G4NDL3.13 library were conducted, and the difference of the respective best fit value was used  as the systematic uncertainty due to the neutron cross section. 
The effect of the count rate difference mentioned in  section \ref{sec:EventSelection} was also evaluated. We evaluated the impact on the scintillation decay time constant by lowering the weight of events in the data below 20 PEs and it turned out to be a negligible effect.
\item (3)L$_{\rm eff}$: following Ref.~\cite{Aprile:2011hi}'s error estimates, we ran MC simulations also  with L$_{\rm eff}$ $\pm$1 $\sigma$. We used the difference of the respective best fit values  as the systematic uncertainty due to L$_{\rm eff}$.

\item (4){\it Jitter}: timing jitter affects  the determination of the rising edge of  timing distributions. The uncertainty was evaluated by comparing the timing distribution of data and simulated samples with  different assumptions for the amount of  timing jitter.   $t_{\rm jitter}$  was changed from $\sigma$ = 0.93 ns to 0.0, 0.5, 1.5 ns and the differences of the respective best fit value were assigned as  systematic uncertainty.

\item (5){\it After-pulse and pre-pulse}: occasionally single-PE pulses are observed prior to or after the main event pulse, these are aptly labeled pre-pulses and after-pulses, respectively. Their rate and timing information was measured independently in a laboratory setup. The pre-pulses were found to have a 0.10\% /PE probability of occurring and on average are located $\sim$15 ns before the main pulse in a time width of $\sim$2 ns.  After-pulses have a probability of 0.65\% /PE and are located $\sim$40 ns after in a time width of $\sim$5 ns. To study their impact, we generated single-PE pulse timings at the appropriate rate using a Gaussian distribution centered at 40 ns with a 5 ns width and another at -15 ns with a 2 ns width for the after-pulses and pre-pulses, respectively. The differences of the best fit values between using the standard MC PE pulse times and these same times augmented with generated pre-pulse and after-pulse times is used for their systematic uncertainty.
%While studying PMTs we found single-PE level after-pulse about 40 ns after the main pulse  with about 5 ns width, and pre-pulse about 15 ns before with about 2 ns width. The probabilities of the after-pulse and pre-pulse  are 0.65\% and 0.10\% for each PE, respectively.  The rate and timing of both the after- and pre-pulse were measured independently in a laboratory setup. These after- and pre- pulse are not included in our standard MC simulation. To study the impact of these pulses, we added the appropriate rates of after- and pre- pulse with Gaussian distributions: timing at +40 ns with  5 ns width for the after-pulse, and  timing at -15 ns, with 2 ns width for the pre-pulse.  The difference between the two evaluation on the standard and the after- and pre-pulse implemented MC simulation is used as  systematic uncertainty. 

 \end{description}

After adding these systematic uncertainties in quadrature,  the scintillation decay time constant for NR was estimated to be ($\tau_{T}$, $F_{S}$) = (26.9$\pm$0.5 (stat)$^{+0.5}_{-1.0}$ (sys) ns , 0.252$\pm$0.013 (stat)$^{+0.024}_{-0.014}$ (sys))  with $\tau_{S}$ = 4.3$\pm$0.6 ns. 
 
\begin{figure}\centering
\includegraphics[width=10cm]{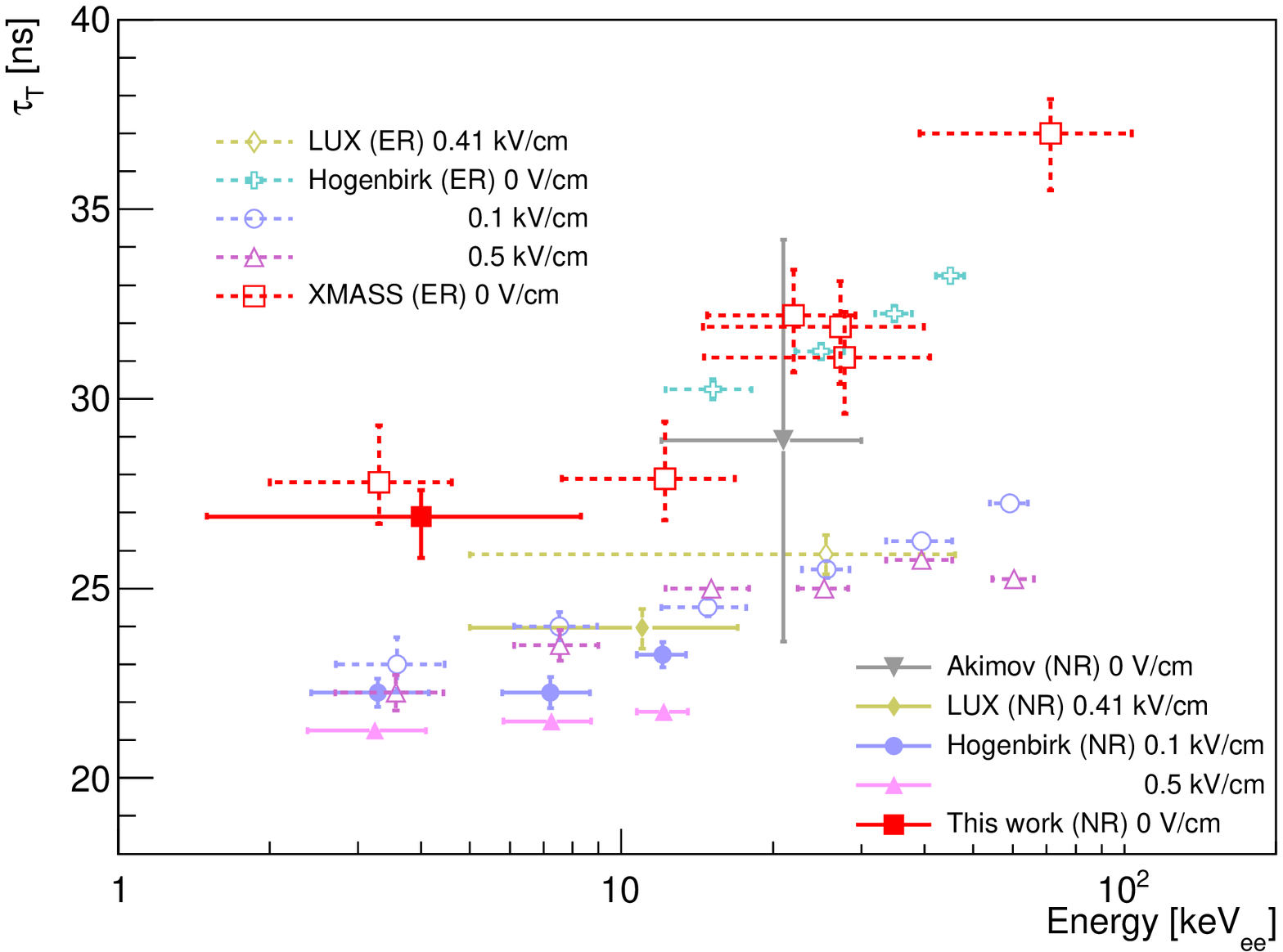}
\includegraphics[width=10cm]{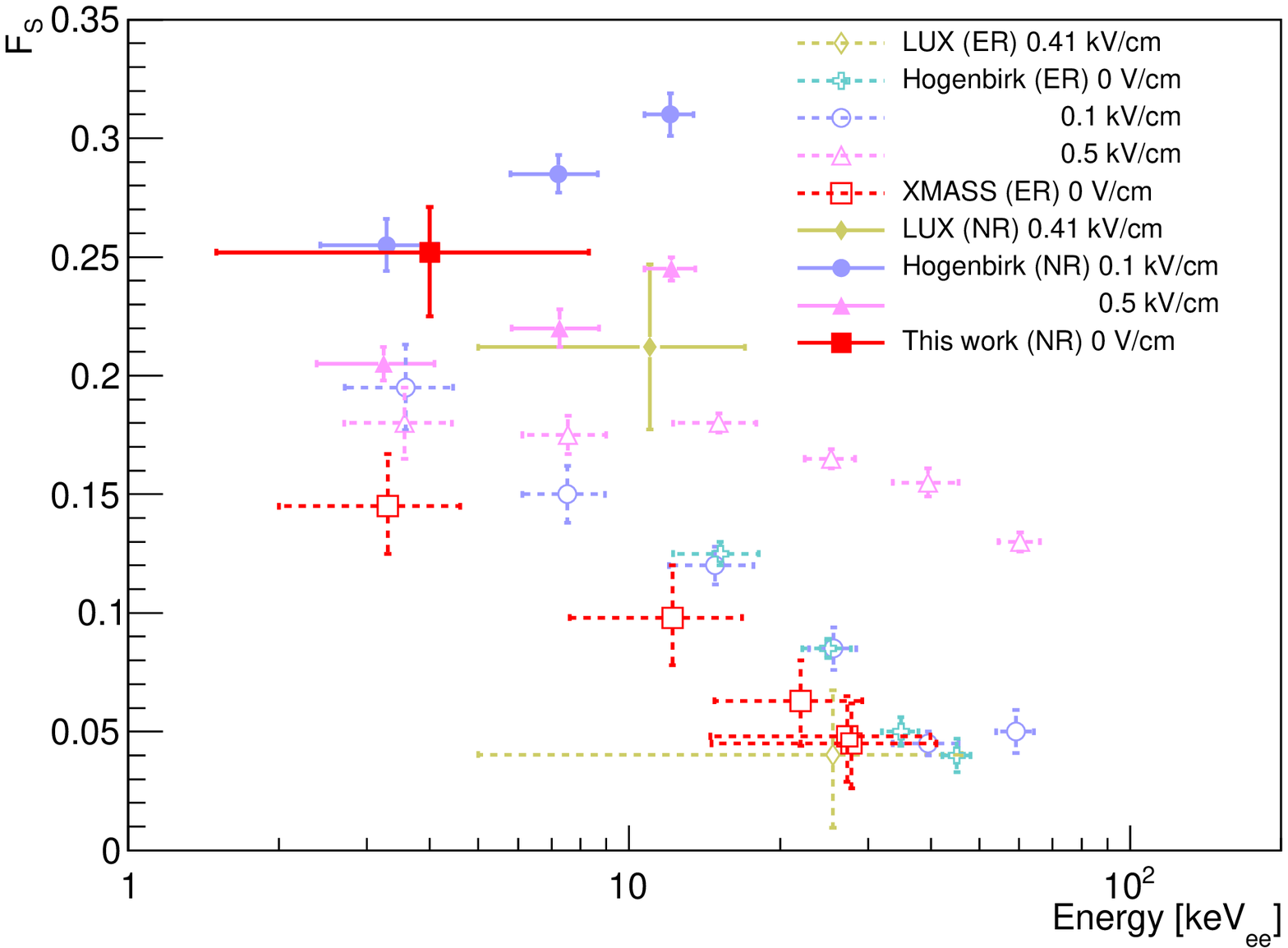}
\caption{ The best-fit parameters $\tau_{T}$ and F$_{S}$ for various measurements. Filled markers and solid lines correspond to NR measurements. Open markers and dotted lines correspond to ER measurements.  Results from Akimov et al. \cite{Akimov:2001pb}, Teymourian et al.  \cite{Teymourian:2011rs}, the LUX experiment  \cite{Akerib:2018kjf}, Hogenbirk et al. (0 V/cm, 0.1 kV/cm and 0.5 kV/cm) \cite{Hogenbirk:2018zwf}, and XMASS measurement (This work and \cite{Takiya:2016wio}) are indicated by the black triangle,  cyan cross, yellow diamond, green cross, blue circle, magenta triangle, and red square, respectively. }
\label{fig:ResultComparison}
\end{figure}

The obtained $\tau_{T}$ is close to that for  ER (27.8$^{+1.5}_{-1.1}$ ns using $^{55}$Fe 5.9 keV gamma-rays ) reported in Ref.~\cite{Takiya:2016wio}, 
although the obtained $F_{S}$ is larger than that of ER (0.145$^{+0.022}_{-0.020}$ using $^{55}$Fe 5.9 keV gamma-ray).
Figure~\ref{fig:ResultComparison} shows  $\tau_{T}$ and F$_{S}$ for various NR and ER measurements.  This measurement had the lowest energy threshold of all the experiments conducted without an external electric field.
D. Akimov {\it et al.} reported a scintillation decay time constants using a single component exponential fit  \cite{Akimov:2001pb}.
 The single component fit value of $\tau$ = 22.5 ns for 1.5 < E < 8.3 keV$_{\rm ee}$ in this work is close to their reported value, although the MC simulation does not reproduce the data well ($\chi^{2}$ / ndf = 368.1 / 116 ).  
 The singlet fraction obtained in this work agrees with the results of Ref.~\cite{Akerib:2018kjf,Hogenbirk:2018zwf}, however, the $\tau_{T}$ is about 5 ns longer than those values. 
 The $\tau_{T}$ discrepancy might stem from a time delay introduced by the recombination process, which is suppressed under an electric field. 
 The recombination process contributes at most 10\% of total scintillation light for nuclear recoil \cite{Aprile:2006kx}. 
 For alpha particles and  $^{252}$Cf fission fragments, this process is thought to be very fast and also have only a minor influence, for these species $\tau_{T}$ was reported as  22$\pm$1.5 ns and 21$\pm$2 ns, respectively~\cite{Hitachi:1983zz} without an applied electric field.
 %In fact, the $\tau_{T}$ for the alpha particle and $^{252}$Cf fission fragments were  22$\pm$1.5 ns and 21$\pm$2 ns, respectively, where the recombination process is very fast and the influence of recombination on the pulse shape is thought to be minor \cite{Hitachi:1983zz}.
  
 %The $\tau_{T}$ is consistent with the value of NR measurement under the electric field.
%As for $\tau_{T}$ for the alpha particle and $^{252}$Cf fission fragments, Hitachi  {\it et al.}  reported 22$\pm$1.5 ns and 21$\pm$2 ns for alpha particle and $^{252}$Cf fission fragment, respectively under the  zero electric field \cite{Hitachi:1983zz}. 
%In their measurement, the recombination process was reported to be very fast and the influence of recombination on the pulse shape was thought to be minor. Their $\tau_{T}$ was consistent with the value reported by Ref. \cite{Akerib:2018kjf,Hogenbirk:2018zwf}, with  neutron source under an electric field. 
%On the other hand,  $\tau_{T}$ in this work is about 5 ns longer than the values measured under the electric field. From this point of view, the difference between $\tau_{T}$ in this work and the result of Ref.  \cite{Akerib:2018kjf,Hogenbirk:2018zwf} might be an effect of the recombination process.
%On the other hand, our $\tau_{T}$ is about 5 ns longer than the values measured under the electric field. This might be an effect of the recombination process.

\subsection{Performance of the pulse shape discrimination}
\begin{figure}\centering
\includegraphics[width=7.5cm]{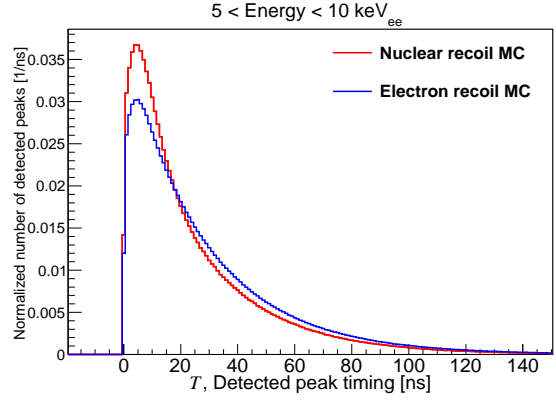}
\caption{Pulse timing distribution of the simulated ER  and NR events with energy deposition from 5 to 10 keV$_{\rm {ee}}$.  In this figure,  TOF was subtracted  and the timing of the fourth earliest peak in each event was shifted to T=0 ns to reflect the trigger implementation. Areas are normalized to 1.}
\label{fig:PSD_PDF}
\end{figure}

\begin{figure}\centering
\includegraphics[width=7.5cm]{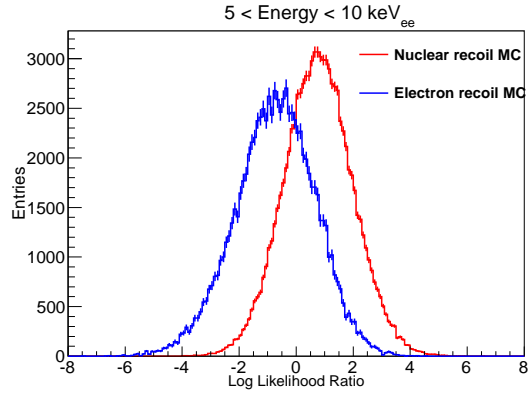}
\caption{Log likelihood ratio distributions of  simulated ER (blue) and NR events (red) reconstructed within a 20 cm radius from the detector center. A positive (negative) log likelihood ratio corresponds to NR (ER) like events. }
\label{fig:LikelihoodRatio}
\end{figure}

For Weakly Interacting Massive Particle (WIMP)  searches in data from a single phase LXe detector, the possibility of PSD between NR and ER is of significant interest. 
We evaluated the performance of  PSD in XMASS-I based on our scintillation decay time constant measurement.  
To obtain the relevant timing distributions, we first simulated the ER events with uniform energy from 0 to 20 keV and  NR events which followed the energy distribution of 100 GeV/c$^{2}$  WIMPs elastically scattering in the LXe target at the center of the detector.
 Figure~\ref{fig:PSD_PDF} shows the peak timing distributions of those simulated ER and NR events that had energy deposits from 5  to 10 keV$_{\rm {ee}}$. 
In Fig.~\ref{fig:PSD_PDF},  TOF was subtracted using the velocity of 110 mm/ns for light in LXe and the timing of the fourth earliest peak in each event was set to $T$ = 0 ns again to reflect the trigger implementation in DAQ.  
As mentioned in section \ref{sec:NRDecayTime},  F$_{S}$ in NR is larger than in  ER. Therefore a  difference in the timing distributions can  clearly be seen.
 These histograms in  Fig.~\ref{fig:PSD_PDF} were used as the probability density function ($f_{\rm ER,NR}(t_{i}$)) of the PMTs hit timings and  we evaluated the following log likelihood ratio
\begin{equation}
\label{eq:LHRatio}
\ln\left(\frac{{L}_{\rm NR}}{{L}_{\rm ER}}\right)  = \sum \ln ( f_{\rm NR}(t_{i})) - \sum \ln (f_{\rm ER}(t_{i}) )~.
\end{equation}
This log likelihood ratio was calculated using PMT hit times  T > 0 ns, after TOF subtraction and T$_{0}$ determination. TOF subtraction uses the reconstructed event position.  The performance of this  PSD method was evaluated using another MC simulation. 
Electron events and NR events where simulated with the same energy distributions as before, but now generated uniformly throughout the detector.  Figure~\ref{fig:LikelihoodRatio} shows the log likelihood ratio distribution of the simulated ER and NR events that have energy deposits from 5 to 10 keV$_{\rm {ee}}$ and were reconstructed within 20 cm from the detector center, these events were used for the evaluation of the PSD performance.

\begin{figure}\centering
\includegraphics[width=7.5cm]{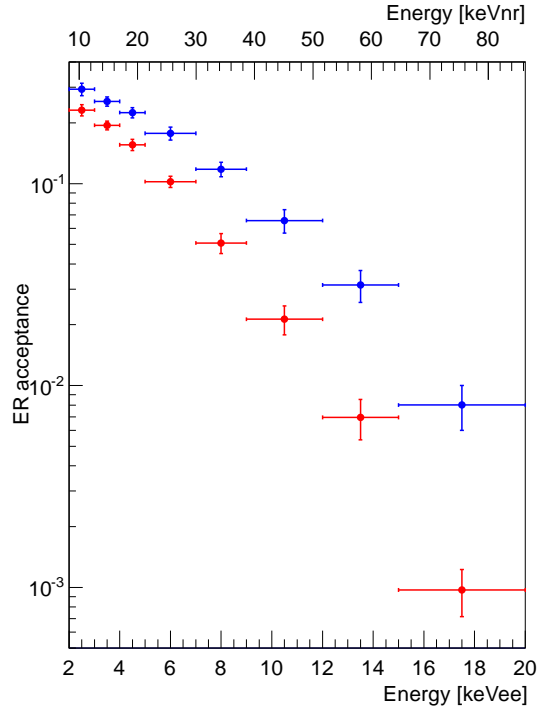}
\caption{ER acceptance as a function of energy. NR equivalent energy is indicated on the top scale. WIMPs with a mass of 100 GeV/c$^{2}$ were used in the NR simulation. Electrons with energy distributed uniformly between 0 and 20 keV were used in the ER simulation. Both MC simulations generated uniformly distributed events within the XMASS-I detector and events which were reconstructed within 20 cm from the detector center were used for the evaluation. The two curves correspond to the log likelihood performance evaluated for a 50\% NR acceptance. Error bars show the quadratic sum of systematic uncertainty and statistical uncertainty. The blue curve shows the result of XMASS-I detector, and the red one represents an ideal case where measurements are not affected by timing jitter in the electronics and TTS in the PMTs.}
\label{fig:PSDResult}
\end{figure}

The ER acceptances when requiring a 50\% NR acceptance for  energies between 5 to 10 keV$_{\rm ee}$ and between 10 to 15 keV$_{\rm ee}$ were estimated to be 13.7${\pm}$1.0\% and 4.1${\pm}$0.7\%, respectively. This corresponds  to a S/$\sqrt{\rm N}$ ratio of  1.4 and 2.5,  respectively. In Fig.~\ref{fig:PSDResult} the blue curve shows the ER acceptance as a function of energy.  The  performance of this PSD method when evaluated using the XMASS-I detector simulation is consistent with the performance that we reported previously using a small chamber \cite{Ueshima:2011ft}.  We also evaluated the performance of this PSD in an ideal case where the measurement is not affected by the timing jitter or TTS (red curve in Fig.~\ref{fig:PSDResult}).  In this ideal case, the PSD performance improves by about a factor of 2 between 5 and  7 keV$_{\rm ee}$, and by about one order of magnitude between 15 and 20 keV$_{\rm ee}$.

\section{Conclusions}
We evaluated the time profile of  NR scintillation emission in LXe  with the XMASS-I detector using  $^{252}$Cf sources.  Two decay components are needed to reproduce the timing distribution of the NR data. We obtained the decay time constant of triplet state $\tau_{T}$ = 26.9$\pm$0.5  (stat)$^{+0.5}_{-1.0}$ (sys) ns and the singlet fraction F$_{S}$ = 0.252$\pm$0.013 (stat)$^{+0.024}_{-0.014}$ (sys) with a decay time constant of singlet state $\tau_{S}$ = 4.3$\pm$0.6 ns taken from a prior research. This measurement had the lowest energy threshold without an aplied  electric field.   We also developed a PSD method based on a log likelihood ratio. The ER acceptances with a 50\% NR acceptance at energies between 5 and 10 keV$_{\rm ee}$ and between 10 and 15 keV$_{\rm ee}$ were estimated to be 13.7${\pm}$1.0\% and 4.1${\pm}$0.7\%, respectively.

\acknowledgments
We gratefully acknowledge the cooperation of the Kamioka Mining and Smelting Company.
This work was supported by the Japanese Ministry of Education,
Culture, Sports, Science and Technology, Grant-in-Aid for Scientific Research, 
JSPS KAKENHI Grant No. 19GS0204 and 26104004,
the joint research program of the Institute for Cosmic Ray Research (ICRR), the University of Tokyo,
and partially by the National Research Foundation of Korea Grant funded
by the Korean Government (NRF-2011-220-C00006), and Institute for Basic Science (IBS-R017-G1-2018-a00).

%\end{linenumbers}

\end{document}